\documentclass[letterpaper,10pt]{article}

\usepackage[utf8]{inputenc}

\usepackage{amsmath,amssymb}

\usepackage{color}
\usepackage{graphicx}
\usepackage{wrapfig}

\usepackage{todonotes}
\usepackage{xspace}
\usepackage{osameet2}
\usepackage{caption}
\graphicspath{{figures/}}

\usepackage[pdftex,colorlinks=true,bookmarks=false,citecolor=blue,urlcolor=blue]{hyperref} %pdflatex
\usepackage{float}

\begin{document}

\title{\text{LDPC Coded Modulation with Probabilistic Shaping} for Optical Fiber Systems}

\author{Tobias Fehenberger\textsuperscript{(1)}, Georg Böcherer\textsuperscript{(1)}, Alex Alvarado\textsuperscript{(2)} and Norbert Hanik\textsuperscript{(1)}}
\address{\textsuperscript{(1)}Institute for Communications Engineering, Technische Universität München, 80333 Munich, Germany\\
 \textsuperscript{(2)}Optical Networks Group, University College London (UCL), London, WC1E 7JE, UK}
\email{tobias.fehenberger@tum.de}

\begin{abstract}
An LDPC coded modulation scheme with probabilistic shaping, optimized interleavers and noniterative demapping is proposed. Full-field simulations show an increase in transmission distance by 8\% compared to uniformly distributed input.
\end{abstract}


%%%%%%%%%%%%%%%%%%%%%%%%%%%%%%%%%%%%%%%%%%%%%%%%%%%%%%%
%              Introduction                           %
%%%%%%%%%%%%%%%%%%%%%%%%%%%%%%%%%%%%%%%%%%%%%%%%%%%%%%%
\section{Introduction}
In the pursuit of higher data rates in optical communications, high-order modulation formats are an established technique to increase spectral efficiencies.
% Possible constellation sizes are in practice often limited by hardware components such as digital converters. Hence, alternatives to switching to the next modulation size have been investigated.
As the optical channel is effectively power-limited by the nonlinearities, optimizing the signaling is necessary in order to further improve the spectral efficiency without increasing the launch power.
Probabilistic shaping and geometric shaping are two possibilities to achieve this. The former uses constellations with nonuniform distributions on a regular grid while the latter uses nonequidistant spacings of equiprobable symbols.
% Constellations with nonuniform distributions on a regular grid (probabilistic shaping) and non-equidistant spacings of equiprobable symbols (geometric shaping)
% \cite{djordjevic2010ipm}
% are two possibilities to achieve this.
% The latter imposes strong requirements on hardware due to the large number of amplitude levels and is not considered here. 
% We expect probabilistic shaping, in contrast, to have less hardware requirements because the signal set stays on an equidistant, i.e. rectangular, grid.\\
The main implementation advantage of probabilistic shaping over geometric shaping is that it does not require modifications of the digital-to-analog
% When comparing geometric and probabilistic shaping as to their complexity, probabilistic shaping has the advantage that analog-to-digital 
converters and the optical signal processing algorithms.
 % do not have to be made aware of it.\par
% For the same spectral efficiency, geometric shaping requires more signal points than probabilistic shaping. This increases the transmitter and receiver complexity 

Probabilistic shaping has been proposed for optical transmission, e.g. in \cite{smith2012trellisshaping,beygi2014polarshaping,yankov2014shaping}.
In \cite{smith2012trellisshaping}, a multi-level coded modulation scheme with trellis shaping and hard-decision decoding is found to operate close to a capacity estimate. Probabilistic shaping via shell-mapping improving a 4D coded modulation scheme is presented in \cite{beygi2014polarshaping}. In \cite{yankov2014shaping}, probabilistic shaping is implemented for turbo codes by a many-to-one mapping. This type of mapping, however, requires information exchange between the demapper and the decoder at the receiver, known as iterative demapping.

In this paper, we adapt the probabilistic shaping scheme introduced in \cite{boecherer2014ShapingLikeARealMan} to a multi-span wavelength-division multiplexing (WDM) optical fiber system with quadrature amplitude modulation (QAM). Off-the-shelf binary encoders and decoders are used and no iterative demapping is required. Using bit-interleaved coded modulation and a low-density parity-check (LDPC) code, the presented scheme increases the transmission distance by 8\%. To the best of our knowledge, this is the first demonstration of such gains based on probabilistic shaping in optical communications. 

%% Shortened
 % and find a \px that works well for the AWGN channel. Instead of extensively computing the optimal distribution using the Blahut-Arimoto algorithm, a less complex heuristic is applied.
% We scale the input constellation by a constant $\Delta$ such that the power constraint imposed by the SNR is fulfilled: $E[|\Delta X|^2]\leq \textrm{SNR}$. For this scaling, we heuristically choose the input distribution that maximizes entropy under an average-power constraint. This is a Maxwell-Boltzmann distribution of the form $p_X(x_i)=K\cdot e^{-|\Delta x_i^2|}$ where $K$ is a normalization constant.
% This optimization problem of reasonable complexity because mutual information can be shown to be unimodal in $\Delta$ for the Maxwell-Boltzmann distribution.\\
% In a last step, it is beneficial to choose the closest (in terms of informational divergence) distribution to \px with entropy equal to the operating point plus $(1-R) \cdot \log_2M$. \todo{Ask Georg how to explain this better and what happens if we dont do this.} This is required because the granularity of the code rates of the LDPC code we use is limited. 
%% / Shortened
% We observe that the system is not sensitive against small changes of \px yet greatly appreciates the rate matching.
% \subsection{Adding redundancy in two stages}\label{ssec:entropy}
{ 
\begin{figure}[b]
	\begin{center}
	\includegraphics{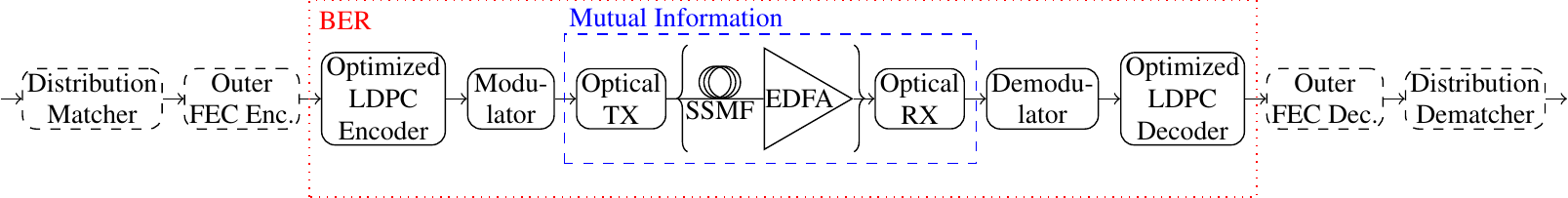}
	\setlength{\abovecaptionskip}{3pt}
	\setlength{\belowcaptionskip}{-5pt}
	\captionof{figure}{\footnotesize Block diagram of the coded modulation system with probabilistic shaping. Each span consists of 100~km SSMF and an EDFA.}
	\label{fig:simsetup}
	\end{center}
\end{figure}
}
% Effectively, redundancy is added in two stages, by shaping and FEC encoding, instead of one stage, the encoder only.
% Suppose the input is 16-QAM and the operating point (the spectral efficiency) is 3 bit/symbol information rate (which is what we use later). The most common approach is to have 1 bit of redundancy for every 3 bit of uniform data, which corresponds to a code rate of 3/4.
%% Shortened
% If the input $X$ consisting of $M$ elements is not uniformly distributed any longer, its entropy decreases, i.e. less information is contained in it. For a uniform distribution, the entropy $H(X)$ equals $\log_2(M)$, whereas for a shaped input, $H(X)$ is smaller because redundancy is contained into the shaped symbols. This becomes important for determining the code rate $R$. Suppose the input is 16-QAM and the operating point (the spectral efficiency) is 3 bit/symbol information rate (which is what we use later). The most common approach is to have 1 bit of redundancy for every 3 bit of uniform data, which corresponds to a code rate of 3/4. As the shaped data already contains redundancy, the code rate must be larger than 3/4 to still operate at 3 bit/symbol. Effectively, redundancy is added in two stages, by shaping and FEC encoding, instead of one stage, the encoder only.\
%% /Shortened
%%%%%%%%%%%%%%%%%%%%%%%%%%%%%%%%%%%%%%%%%%%%%%%%%%%%%%%
%              Implementation                         %
%%%%%%%%%%%%%%%%%%%%%%%%%%%%%%%%%%%%%%%%%%%%%%%%%%%%%%%
\section{Mutual Information Analysis}\label{sec:MIsim}
\subsection{Optical Transmission Setup}\label{ssec:implementation}
% We evaluate symbolwise MI for an optical system as shown in the dashed blue box in Fig.~\ref{fig:simsetup}.
The setup of the optical coded modulation system is shown in Fig.~\ref{fig:simsetup}. The QAM symbols are either uniformly distributed or shaped as explained in Sec.~\ref{ssec:shaping}. We consider a dual-polarization multi-span WDM system with 15 co-propagating channels. The baud rate per channel is 28 GBaud, the pulses are root-raised-cosine (RRC) shaped with 5\% roll-off, the WDM spacing is 30~GHz. We simulate 2\textsuperscript{16} symbols per polarization. The fiber is a standard single-mode fiber (SSMF) with $\alpha$=0.2~dB/km, $\gamma$=1.3 (W$\cdot$km)\textsuperscript{-1} and $D$=17~ps/nm/km. Each span of length 100~km is followed by an Erbium-doped fiber amplifier (EDFA) with a noise figure of 4~dB. Signal propagation is simulated using the split-step Fourier method with 32 samples per symbol and a step size of 100~m. At the receiver, the center WDM channel is bandpass-filtered and converted into the digital domain with a coherent receiver. Chromatic dispersion is compensated digitally and an RRC filter is applied again. Laser phase noise and polarization mode dispersion are not included in the simulations as perfect compensation is assumed. We compute nonparametric estimates of the mutual information (MI) between input and output symbols without considering memory (see the dashed blue box in Fig.~\ref{fig:simsetup}). In particular, Gaussian statistics are not assumed a priori for MI estimation.
% Ideal data-aided equalization is performed to jointly rotate the received constellation to its correct orientation in the complex plane.
% After an RRC filter, MI is estimated.

%%%%%%%%%%%%%%%%%%%%%%%%%%%%%%%%%%%%%%%%%%%%%%%%%%%%%%%
%        Probabilistic Constellation Shaping          %
%%%%%%%%%%%%%%%%%%%%%%%%%%%%%%%%%%%%%%%%%%%%%%%%%%%%%%%
\subsection{Optimized Input Distribution}\label{ssec:shaping}
% \subsection{How to find the nonuniform input distribution \px}\label{ssec:howtofindpx}
% We use the following approach to find the shaped distribution \px for the input $X$.
We probabilistically shape 16-QAM and 64-QAM by assigning larger probabilities to the points with lower energy. At the same power, this increases the Euclidean distance between constellation points compared to uniform input.

To find a suitable distribution for the shaped input, we first resort to the Gaussian noise (GN) model \cite{poggiolini2014GN} and calculate the signal-to-noise ratio (SNR) for every transmission distance. We then follow \cite{kschischang1993mbdistribution} and jointly optimize the Maxwell-Boltzmann distribution of the input symbols and the constellation scaling such that the MI is maximized for the additive white Gaussian noise (AWGN) channel under the power constraint imposed by the SNR.
The obtained distribution is used for the shaped input of the optical fiber simulations.

Insets b) and c) of Fig.~\ref{fig:MIvsL} illustrate the effect of uniform and shaped input on the received symbol distributions, respectively, with hotter colors representing a larger number of occurrences. The innermost points of the shaped input are sent more often and the Euclidean distance between constellation points is 18\% larger than for uniform input.
\subsection{Mutual Information Results}
\begin{figure}[t]
	\centering
	\includegraphics{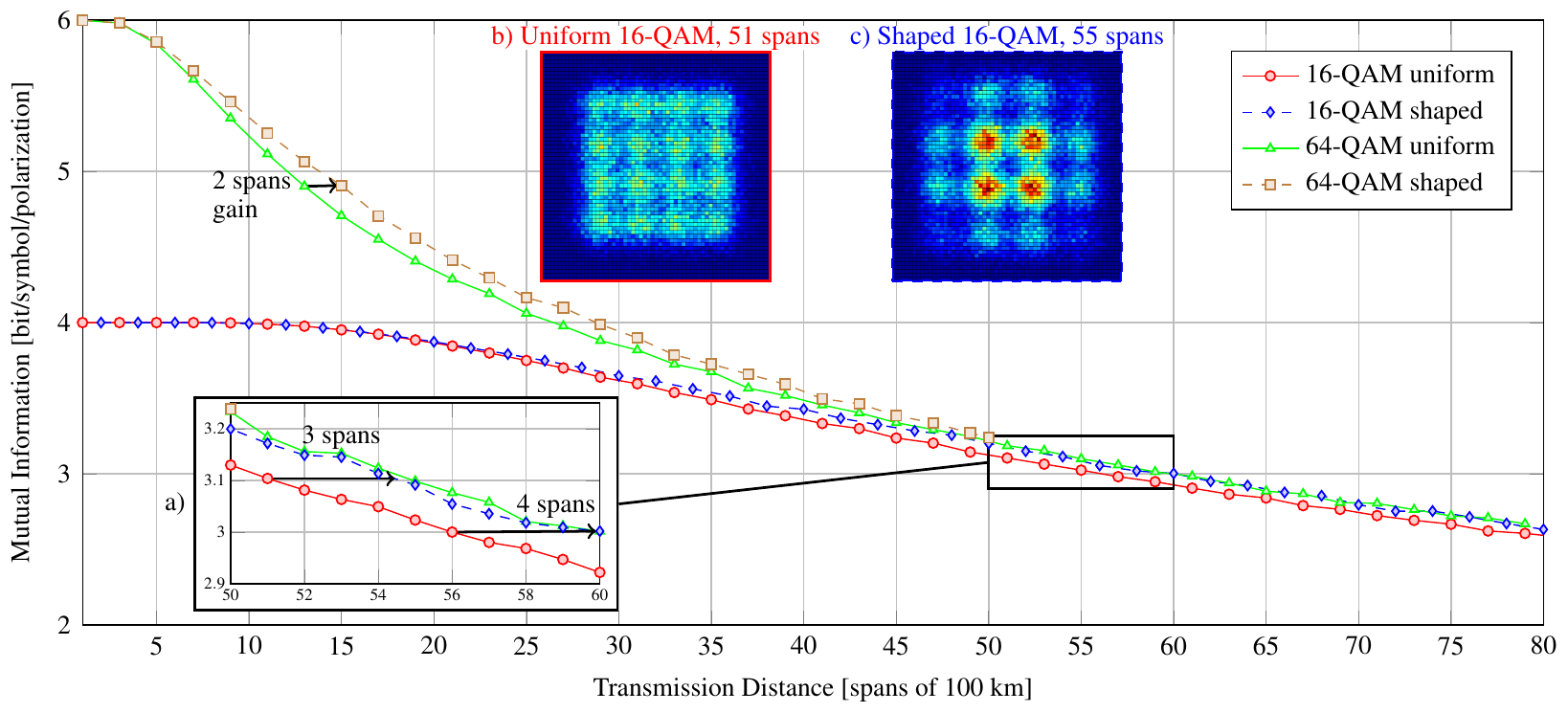}
	\setlength{\abovecaptionskip}{-5pt}
	\setlength{\belowcaptionskip}{-10pt}
\captionof{figure}{\footnotesize MI versus distance. a): Zoom from 50 to 60 spans. b) and c): Distributions of the received symbols for uniform and shaped input, resp.}
\label{fig:MIvsL}
\end{figure}

MI versus transmission distance is shown in Fig.~\ref{fig:MIvsL} for 16-QAM and 64-QAM, both with uniform and shaped input. For each overall transmission distance of 1 to 80 spans, we use an optimized shaped input and determine a candidate for the optimal launch power from the GN model. By testing different launch powers in our simulations, we found that the optimum is within 0.2~dB of $-1.6$~dBm for all simulated modulations and distances.
% For the shaped curve, the input distribution is chosen as described in Section~\ref{sec:shaping} for every distance.

For small distances, shaped and uniform QAM achieve similar MI because the impact of distortions is small in this region and the shaped distribution is close to the uniform distribution.
% Higher spectral efficiencies in this range demand higher-order QAM such as 64-QAM. 
Using shaped 64-QAM at 15 spans, however, gives 2 additional spans compared to uniform input, which is an increase of 15\%. For longer distances, the improvement over uniform 64-QAM disappears, and thus, we did not simulate shaped 64-QAM for more than 50 spans.
% We do not consider shaped 64-QAM in the following analysis.
% In the following analysis, we focus on longer distances and do not consider shaped 64-QAM.
Once the signal propagates further and distortions accumulate, we observe that shaped 16-QAM performs better than uniform 16-QAM.
%% Add Fig.3 around here to have it at the upper left corner of page 3
% At 25 spans, the MI gap between shaped 16-QAM and uniform 16-QAM starts to open. 
From 50 spans onwards, the MI curves of shaped 16-QAM and uniform 64-QAM coincide.
% increases MI by almost 0.1~bit/symbol for the same distance (vertical direction) and
Inset~a) of Fig.~\ref{fig:MIvsL} shows that shaped 16-QAM increases the transmission distance by 3 to 4 spans compared to uniform 16-QAM.
% 
% points out that shaped 16-QAM increases the distance by 3 to 4 spans for identical MI.
% Results for the AWGN channel suggest that shaped 16-QAM beats uniform 64-QAM in a certain SNR range. We attribute not seeing this behavior to the suboptimality of \px for the optical channel as well as the stronger impact of nonlinearities on an expanded constellation.
%%%%%%%%%%%%%%%%%%%%%%%%%%%%%%%%%%%%
% BER Implementation
%%%%%%%%%%%%%%%%%%%%%%%%%%%%%%%%%%%%
\section{BER Analysis}\label{sec:implementation}
To verify our MI results, we implement an LDPC code as inner forward error correction (FEC) and compare the BERs of uniform and shaped 16-QAM (see the dotted red box in Fig.~\ref{fig:simsetup}). We operate at an information rate of 3 bits per symbol per polarization, without considering the coding overhead (OH) of the outer FEC.
The LDPC codes of our coded modulation scheme are from the DVB-S2 standard and have a block length of 64800 bits.% which will allows us to come close to this asymptotic limit.

% The LDPC codes we use are from the DVB-S2 standard and have a block length of 64800 bits. Using these finite-length codes, we will come within 5 spans of this asymptotic limit.
% The shaped 16-QAM constellation has the probabilities $\{0.025, 0.054, 0.117\}$ for the inner, middle, and outer ring, respectively. This distribution is obtained for a fiber length of 55~spans as described in Section \ref{sec:shaping}. The reason that we cannot transmit over 60 spans as an MI of 3 bit/symbol would suggest (c.f. Fig.~\ref{fig:MIvsL}) but 5 spans less is because the DVB-S2 LDPC codes are not capacity-achieving.
\subsection{Implementation}
 A total of four bits map to one 16-QAM symbol. For uniform input and a target information rate of 3~bits per symbol, we use 3~bits for data and one bit for redundancy. The rate 3/4 LDPC code is used to achieve this. For our shaped input% at 55 spans
, we effectively map 3~uniformly distributed data bits to 3.2~shaped bits. We use the remaining $4-3.2=0.8$~bits per QAM symbol for redundancy by encoding the shaped data with a rate $3.2/4=4/5$ LDPC code. Hence, the overall redundancy of one bit is successively added in two steps, i.e., by probabilistic shaping and by LDPC encoding. 
% The input distribution is made available to the LDPC encoder as priors.
\setlength{\intextsep}{1pt} % Vertical space above & below [h] floats
\begin{wrapfigure}{r}{0.5\textwidth}
	% \vspace{3pt}
	% \includegraphics{BERvsL}
	\includegraphics{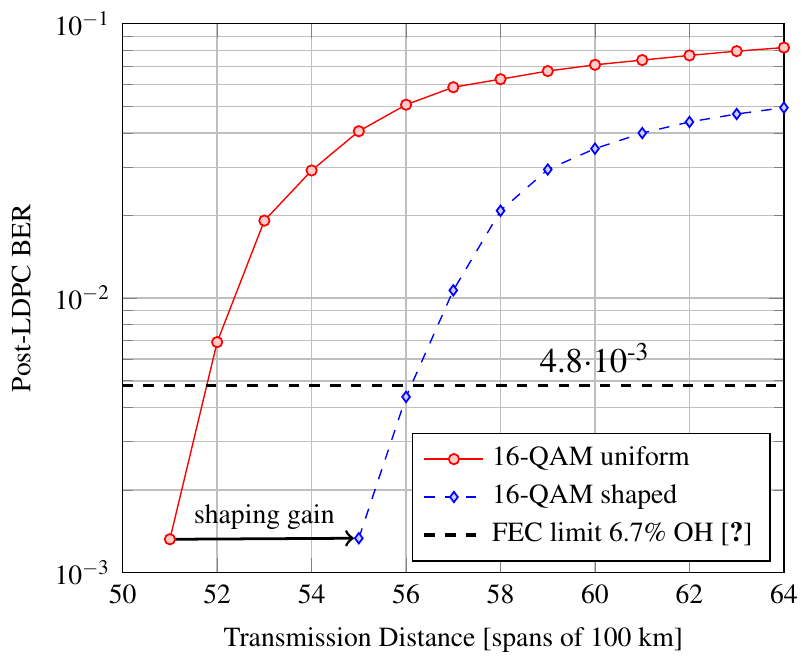}
	\setlength{\abovecaptionskip}{-10pt}
	\setlength{\belowcaptionskip}{-5pt}
	\captionof{figure}{\footnotesize BER after LDPC decoding at 3 bits per symbol information rate.}
	\label{fig:BERvsL}
\end{wrapfigure}
The mapping from uniformly distributed data bits to shaped bits is done by a distribution matcher \cite{baur2014ADM} as shown in Fig.~\ref{fig:simsetup}. In our simulations, we emulated the distribution matcher output by directly generating the shaped bits.\par

We optimize the LDPC codes by wrapping the encoder with two interleavers and the decoder with the respective inverse interleavers. The interleavers optimize the mapping of the coded bits to the QAM symbols and ensure that the distribution imposed on the data is preserved at the modulator output \cite[Sec. V-D]{boecherer2014ShapingLikeARealMan}.\par

A staircase code with 6.7\% OH \cite{smith2012staircase} is assumed as outer code. It must be placed between the distribution matcher and the inner FEC to avoid error propagation. A systematic encoder must be used to preserve the statistics generated by the distribution matcher.

\subsection{BER Results}
In Fig.~\ref{fig:BERvsL}, the plot of BER versus distance shows that the shaping gain predicted by the MI in Fig.~\ref{fig:MIvsL} translates into an improvement in BER.
% Both the uniform and the nonuniform input operate at a rate of 3 bit/symbol.
At a BER of 1.3$\cdot$10\textsuperscript{-3}, we observe a shaping gain of 4~spans compared to uniform 16-QAM. This increase in transmission distance by 8\% is in excellent agreement with the prediction of the MI curves in Fig.~\ref{fig:MIvsL}.

% Fig.~\ref{fig:MIvsL} suggests that at 3 bits per symbol, 60 spans can be achieved with shaped 16-QAM. Using the finite-length LDPC codes, we operate within approximately 5 spans of this asymptotic limit.
%%%%%%%%%%%%%%%%%%%%%%%%%%%%%%%%%%%%%%%%%%%%%%%%%%%%%%%
%              Conclusion                             %
%%%%%%%%%%%%%%%%%%%%%%%%%%%%%%%%%%%%%%%%%%%%%%%%%%%%%%%
\section{Conclusions}
We have presented an LDPC coded modulation scheme with probabilistic shaping. Compared to previously suggested shaping schemes, the implementation has lower complexity because no iterative demapping is required at the receiver. We show in optical fiber simulations that 16-QAM and 64-QAM with shaped input allow for an increase in transmission distance by 8\% and 15\%, respectively, compared to uniformly distributed input.
% We also find that shaped 16-QAM performs as good as uniform 64-QAM in the considered range.
% Finally, we report an increase in distance by 8\%. We verify this by BER simulations using an LDPC code.
The results are promising and further gains are expected by optimizing the input distribution for the nonlinear optical channel. Also, four-dimensional constellations that are more robust against nonlinear distortions will be considered in future work.
%%%%%%%%%%%%%%%%%%%%%%%%%%%%%%%%%%%%%%%%%%%%%%%%%%%%%%%
%              References                             %
%%%%%%%%%%%%%%%%%%%%%%%%%%%%%%%%%%%%%%%%%%%%%%%%%%%%%%%

\end{document}